\documentclass[reprint,
superscriptaddress,
amsmath,amssymb,
]{revtex4-1}

\usepackage{xcolor}
\usepackage{graphicx}
\usepackage{dcolumn}
\usepackage{bm}
\usepackage[utf8]{inputenc}
\usepackage{upgreek}
\usepackage{appendix}

\begin{document}

\title{Exact Analytical Solution of the Flory-Huggins Model and Extensions to Multicomponent Systems}

\author{J. Pedro de Souza}
\affiliation{%
 Omenn-Darling Bioengineering Institute, Princeton University, Princeton, NJ, 08544
}%

\author{Howard A. Stone}
\affiliation{%
 Department of Mechanical and Aerospace Engineering, Princeton University, Princeton, NJ, 08544
}%

\date{\today}

\begin{abstract}

The Flory-Huggins theory describes the phase separation of solutions containing polymers. Although it finds widespread application from polymer physics to materials science to biology, the  concentrations that coexist in separate phases at equilibrium have not been determined analytically, and numerical techniques are required that restrict the theory's ease of application. In this work, we derive an implicit analytical solution to the Flory-Huggins theory of one polymer in a solvent by applying a procedure that we call the implicit substitution method. While the solutions are implicit and in the form of composite variables, they can be mapped explicitly  to a phase diagram in composition space. We apply the same formalism to multicomponent polymeric systems, where we find analytical solutions for polydisperse mixtures of polymers of one type. Finally, while complete analytical solutions are not possible for arbitrary mixtures, we propose computationally efficient strategies to map out coexistence curves for systems with many components of different polymer types.

\end{abstract}

\maketitle

%
%

\section{Introduction}

Since its introduction in the early 1940s first by Huggins~\cite{huggins1941solutions} and then by Flory~\cite{flory1942thermodynamics}, the Flory-Huggins theory has become the most widespread model of phase separation of polymer solutions. Due to its simplicity in construction, the theory, with a minimal set of parameters, captures the main trends in the polymer partitioning between phases.  The theory has been applied in numerous contexts including in chemical processing ~\cite{cantow2013polymer, barton2018handbook}, materials science~\cite{milczewska2006use, peinemann2007asymmetric, angelopoulou2023high, callaway2018molecular}, and also in the burgeoning field of biomolecular condensation~\cite{brangwynne2015polymer, berry2018physical, pappu2023phase}. While many corrections and modifications have been proposed to capture more detailed physical phenomena~\cite{overbeek1957phase, simha1969statistical, scheutjens1980statistical, tanaka1989theory, tanaka1994thermoreversible, semenov1998thermoreversible, ruzette2001simple, wang2008effects, wang201750th, wessen2021simple}, the original Flory-Huggins theory is the standard starting point for any analysis of materials that include polymeric components.

From its physics-based construction, the Flory-Huggins theory outputs a system of nonlinear constraints specifying thermodynamic equilibria that define the coexisting phase concentrations. While the equilibrium conditions can be solved numerically, up to this point, to our knowledge, no exact analytical solution has been found. Therefore, any evaluation of the theory or fitting of the theory to data requires numerical solutions, hampering the ease of use of the theory. In some regimes, analytical approximations can be applied~\cite{sanchez1984corresponding,tsenoglou2001slightly,scheinhardt2004stationary,qian2022analytical, van2023predicting}, but these approximations may involve complicated expressions and may not be uniformly valid, even for a single polymer-solvent combination.

The technical challenge of describing coexisting phases is more apparent for multicomponent mixtures~\cite{mao2019phase, jacobs2017phase, jacobs2023theory}, even for mixtures of the same polymer type with different polymer lengths~\cite{huggins1967theoretical}. In order to construct a phase diagram using direct numerical solution, it is necessary to discretize the compositional space, and solve nonlinear systems of equations at each point. As the number of polymer types, $M$, increases, the discretized space increases exponentially, with a power proportional to $M$, which corresponds to a large number of discrete compositions, especially if fine resolution is desired. In many applications, such as in cellular biology, the number of polymer types can exceed tens of thousands~\cite{wuhr2014deep}. Such a brute force numerical approach may not be feasible to describe the breadth of coexisting concentrations over the full compositional space for $M>4$, never mind for $M>10^4$. 

\begin{figure*}[t!]
\centerline{\includegraphics[scale=0.65]{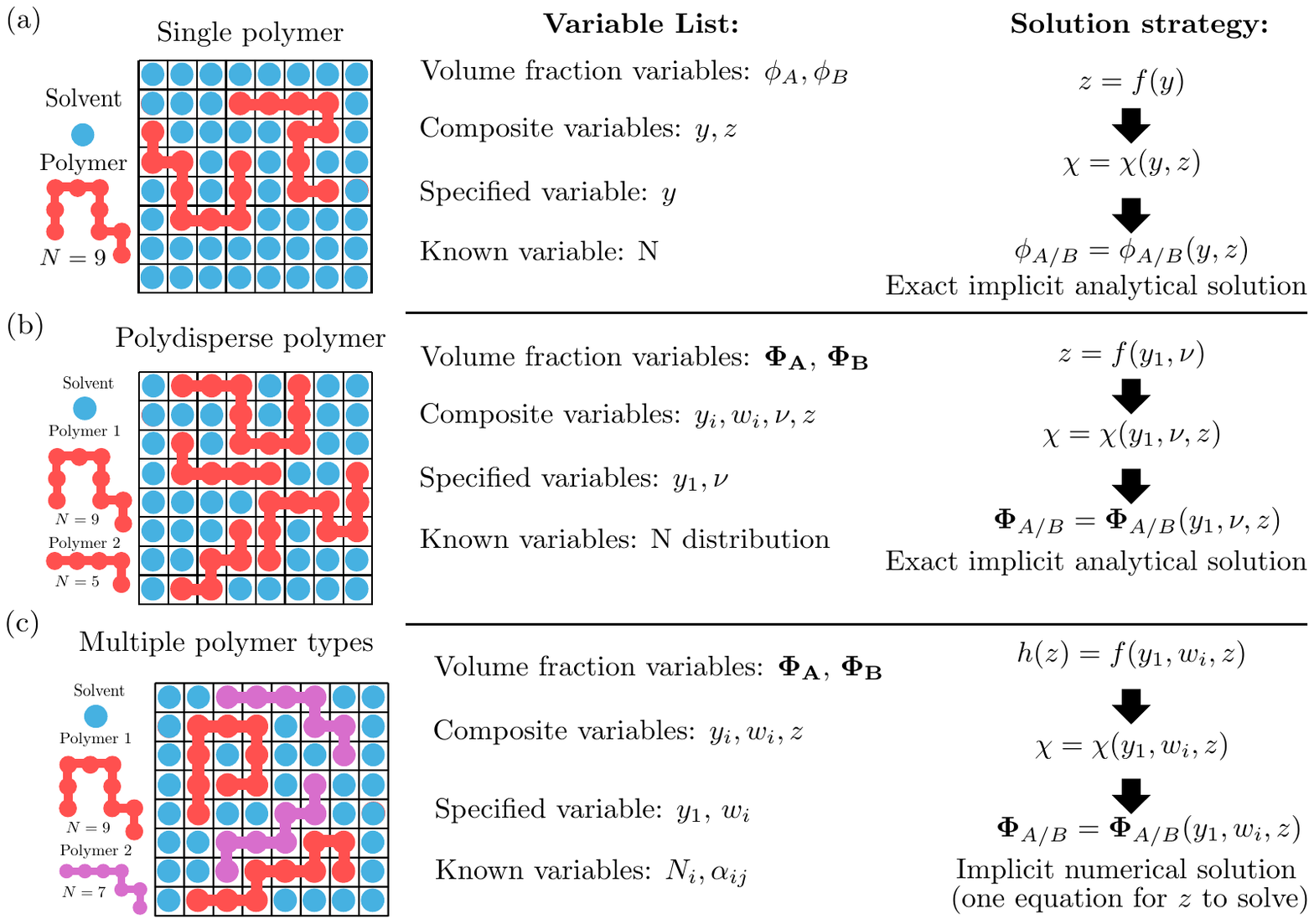}}
\caption{Schematic of the systems under study, their corresponding variables, and the solution strategy employed in each case for (a) a single polymer-solvent solution, (b) a polydisperse polymer-solvent solution, and (c) a solution containing multiple polymer types in solvent.}
\label{Figure1-IllustrationOfMethod}
\end{figure*}

Here, we analytically derive an exact implicit solution to the Flory-Huggins theory of a one polymer-one solvent system using an implicit substitution method. The key idea is to solve for each constraint for the Flory-Huggins $\chi$ parameter, and then substitute the value of $\chi$ into each equation so as to eliminate it from the system. We explore the solution's asymptotic behavior, referring back to the original applications of the theory by Flory~\cite{flory1942thermodynamics}.  The general idea of our approach for this and other problems described in this paper is illustrated in Figure~\ref{Figure1-IllustrationOfMethod}.

Next, we apply the implicit substitution method to multicomponent mixtures where the relative magnitude of the pair interactions are known. There, we discover an exact analytical solution for polydisperse polymer samples of the same type, and distill the phase diagram of these mixtures into two independent variables that fully specify the coexistence curves. Using this strategy, we can analytically describe the composition (molecular weight distribution and total polymer volume fraction) of the coexisting phases.

For generalized mixtures, we use the implicit substitution method to simplify a nonlinear system of $M+1$ equations down to one nonlinear ``master'' equation with one unknown--- the solvent partitioning between the phases. While this reduction leads to great simplification of the equilibrium calculation, in the case of arbitrary mixtures, the master equation must be solved numerically. If desired, we can discretize the compositional space and construct a $\chi$ surface by solving just this one equation at each point---and we demonstrate this idea for a two polymer, one solvent mixture. To minimize the number of function evaluations, we propose a sampling method for the implicit function that obviates the need to discretize the entire compositional space. Choosing sampling points only requires finding the poles of the implicit $\chi$ function at a fixed distance from the global critical point, which we demonstrate, although we leave a full computational implementation of the compositional sampling for a large number of components to future work.

The implicit substitution method developed in this work is a powerful technique to solve or simplify the nonlinear systems of equations encountered in the thermodynamics of solutions. The only price to be paid is that the derived solutions are implicit in composite composition variables, themselves a function that is a combination of the coexisting phase concentrations. Nevertheless, these composite composition variables are experimentally accessible, and we expect the technique will find applications in thermodynamic analyses in various contexts.




\section{The Flory-Huggins Model: One Polymer}
\label{OneComponentCase}
\subsection{Theoretical background}
The Flory-Huggins model may be derived using a free energy density where  a polymer composed of $N$ monomers occupies a lattice. The free energy density per lattice site, $f$, may be expressed in dimensionless terms, $\tilde{f}$, as:
\begin{equation}\label{eq:free_energy_one_polymer}
\begin{split}
  \tilde{f}=\frac{fv}{k_BT}=&\frac{1}{N}\phi\ln(\phi)+(1-\phi)\ln(1-\phi)\\ &+\chi\phi(1-\phi).
  \end{split}
\end{equation}
Here, $v$ is a lattice site volume, $k_B$ is the Boltzmann constant, $T$ is the absolute temperature,  and $\phi$ is the polymer volume fraction. $\chi$ is the so-called Flory parameter, which captures the interaction between solvent and the polymer~\cite{doi2013soft}.

From this free energy, we can define the dimensionless chemical potential of the polymer, ${\mu}=\partial \tilde{f}/\partial \phi$, up to an arbitrary constant,
\begin{equation}
{\mu}=\frac{1}{N}\ln(\phi)-\ln(1-\phi)-2\chi \phi,
\end{equation}
and the dimensionless osmotic pressure of the solution is $\tilde{\Pi}=-\tilde{f}+\phi\tilde{f}^\prime$, so that
\begin{equation}
   {\Pi}=\left(\frac{1}{N}-1\right)\phi-\ln(1-\phi)-\chi\phi^2.
\end{equation}

The model predicts two coexisting phases, a dense phase and a dilute phase of the polymer, above a critical $\chi$ value, $\chi>\chi_c$,
\begin{equation}
    \chi_c=\frac{1}{2}\left(1+\frac{1}{\sqrt{N}}\right)^2
\end{equation}
and beginning at a critical polymer volume fraction $\phi =\phi_c$:
\begin{equation}\label{eq:eqCriticalPoint}
    \phi_c=\frac{1}{1+\sqrt{N}}.
\end{equation}
It is found that as $N$ increases, the phase diagram becomes asymmetric relative to the center line at $\phi=1/2$. 

Chemical equilibrium between two phases $A$ and $B$ at volume fractions $\phi_A$ and $\phi_B$ is specified by two contraints---equal chemical potential of the polymer and equal osmotic pressure of the solvent in the two phases:
\begin{subequations}
    \begin{equation}
        \mu(\phi_A)=\mu(\phi_B)
    \end{equation}
    \begin{equation}
        \Pi(\phi_A)=\Pi(\phi_B).
    \end{equation}
\end{subequations}

These coupled equations can be rewritten explicitly as:
\begin{subequations}
    \begin{equation}\label{eq:eqChemEqA}
        \frac{1}{N}\ln\left(\frac{\phi_A}{\phi_B}\right)-\ln\left(\frac{1-\phi_A}{1-\phi_B}\right)-2\chi (\phi_A-\phi_B)=0 
    \end{equation}
    \begin{equation}\label{eq:eqChemEqB}
        \left(\frac{1}{N}-1\right)(\phi_A-\phi_B)-\ln\left(\frac{1-\phi_A}{1-\phi_B}\right)-\chi(\phi_A^2-\phi_B^2)=0.
    \end{equation}
\end{subequations}

If $\chi$ is known, then we have two equations with two unknowns, $\phi_A$ and $\phi_B$. To our knowledge, these coupled equations are impossible to deconvolve into two explicit functions describing $\phi_A$ and $\phi_B$ with known analytical techniques. However, in order to make analytical progress, we propose two simultaneous strategies. First, we search for implicit solutions, where we treat $\chi$ as an unknown, so as to gain an additional degree of freedom in the compositional space; also, we can eliminate $\chi$ from the system of equations quite easily. Second, instead of working with the variables $\phi_A$ and $\phi_B$, we use two orthogonal composite variables that combine $\phi_A$ and $\phi_B$, which we call $y$ and $z$, that become  apparent after eliminating $\chi$. Thus, the problem is reduced to solving for one unknown variable, $z$, in terms of one known variable $y$. We call this solution method an implicit substitution method, since we use substitution to eliminate the implicit function $\chi$. Now that the strategy has been explained, we implement it in detail below.

\subsection{Analytical solution}
First, we solve the equations (\ref{eq:eqChemEqA}) and (\ref{eq:eqChemEqB}) for $\chi$:
\begin{subequations}
\begin{equation}\label{eq:eqChemEqChiA}
    \chi=\frac{\frac{1}{N}\ln\left(\frac{\phi_A}{\phi_B}\right)-\ln\left(\frac{1-\phi_A}{1-\phi_B}\right)}{2(\phi_A-\phi_B)} 
    \end{equation}
    \begin{equation}\label{eq:eqChemEqChiB}
        \chi=\frac{\left(\frac{1}{N}-1\right)(\phi_A-\phi_B)-\ln\left(\frac{1-\phi_A}{1-\phi_B}\right)}{\phi_A^2-\phi_B^2}.
    \end{equation}
\end{subequations}
We equate these expressions to eliminate $\chi$ and multiply through by a factor $2(\phi_A^2-\phi_B^2)$ to arrive at
\begin{equation}
\begin{split}
        &\left[\frac{1}{N}\ln\left(\frac{\phi_A}{\phi_B}\right)-\ln\left(\frac{1-\phi_A}{1-\phi_B}\right)\right](\phi_A+\phi_B)=\\&2\left(\frac{1}{N}-1\right)(\phi_A-\phi_B)-2\ln\left(\frac{1-\phi_A}{1-\phi_B}\right).
        \label{OnePolymerIntermediateEquation}
\end{split}
\end{equation}
Organizing the terms, we can write (\ref{OnePolymerIntermediateEquation}) as:
\begin{equation}
\begin{split}
        &\frac{\left (\phi_A+\phi_B\right )}{N}\ln\left(\frac{\phi_A}{\phi_B}\right)+(2-\phi_A-\phi_B)\ln\left(\frac{1-\phi_A}{1-\phi_B}\right)=\\&2\left(\frac{1}{N}-1\right)(\phi_A-\phi_B),
\end{split}
\end{equation}
or
\begin{equation}
\begin{split}
        &\frac{\left (\phi_A+\phi_B\right )}{N(\phi_A-\phi_B)}\ln\left(\frac{\phi_A}{\phi_B}\right)+\frac{2-\phi_A-\phi_B}{\phi_A-\phi_B}\ln\left(\frac{1-\phi_A}{1-\phi_B}\right)=\\&2\left(\frac{1}{N}-1\right).
        \label{OnePolymerIntermediateEquation3}
\end{split}
\end{equation}
If we define variables $y$ and $z$ as:
\begin{equation}
\begin{split}
        &y=\frac{\phi_A-\phi_B}{\phi_A+\phi_B}\\
        &z=\frac{\phi_A-\phi_B}{2-\phi_A-\phi_B}
\end{split}
\end{equation}
equation (\ref{OnePolymerIntermediateEquation3}) becomes:
\begin{equation}
    \frac{1}{Ny}\ln\left(\frac{1+y}{1-y}\right)-\frac{1}{z}\ln\left(\frac{1+z}{1-z}\right)=2\left(\frac{1}{N}-1\right).
    \label{OnePolymerIntermediateEquation5}
\end{equation}
The variables $y$ and $z$ are relative measures of the partitioning of polymer and solvent, respectively, between phases, and they are already commonly used as an order parameter that defines the phase composition differences going back at least to Cahn and Hilliard~\cite{cahn1958free}. The variable $z$ is similar to $y$ in that it contains the difference over the sum of solvent in each phase,
\begin{equation}\label{eq:eqZsolvent}
    z=\frac{\phi_{sB}-\phi_{sA}}{\phi_{sA}+\phi_{sB}},
\end{equation}
where $\phi_s=1-\phi$ is the solvent volume fraction by incompressibility. It should be noted, however, that these mappings to $y$ and $z$ are not conformal. Nevertheless, since we have one degree of freedom, we will specify the variable $y$, and then solve the above equation for $z$.

In this pursuit, we first can recognize the $\tanh^{-1}$ function, defined by the identity
\begin{equation}
    \tanh^{-1}(y)=\frac{1}{2}\ln\left(\frac{1+y}{1-y}\right),
\end{equation}
to rewrite equation (\ref{OnePolymerIntermediateEquation5}) as
\begin{equation}\label{eq:eqatanh}
    \frac{\tanh^{-1}(y)}{Ny}-\frac{\tanh^{-1}(z)}{z}=\frac{1}{N}-1.
\end{equation}


\begin{figure}[t]
\centerline{\includegraphics[scale=0.45]{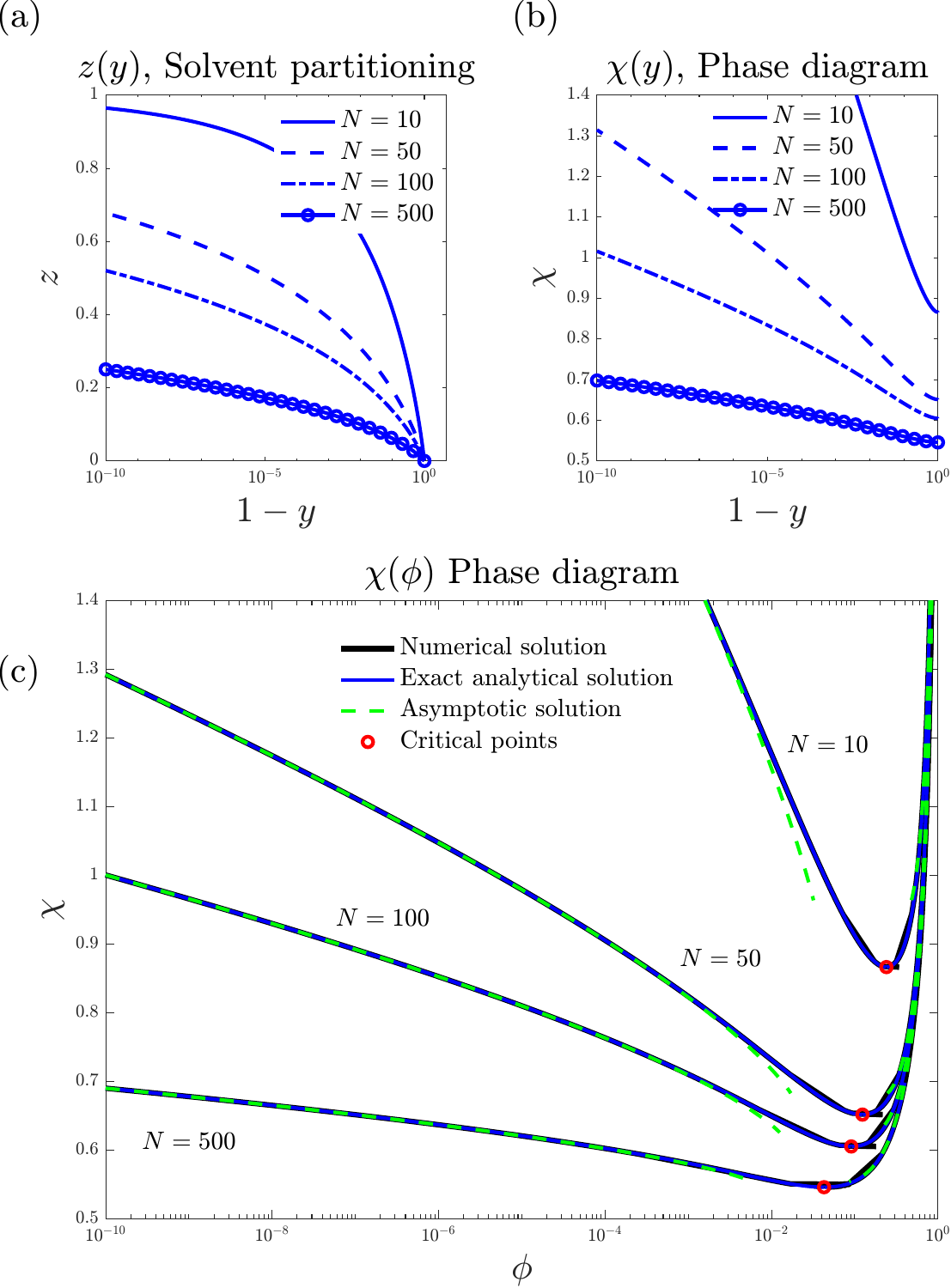}}
\caption{Polymer-solvent solution. Phase diagram for a single polymer-solvent solution,  comparing different degrees of polymerization ($N=10, 50, 100, 500$). (a) Shows the analytical function $z(y)$ that describes the solvent partitioning as a function of the polymer partitioning between phases, (b) shows the analytical function $\chi(y)$, and  (c) maps the coexistence curve to $\chi(\phi)$. The numerical solution overlaps perfectly with the exact analytical solution derived here. For large $N$, the asymptotic approximation first used by Flory~\cite{flory1942thermodynamics} matches most of the curve far from the critical point. The red circles mark the critical points at each value of $N$.}
\label{Figure2PolymerSolventSolution}
\end{figure}

If we define the function $h(x)$ as
\begin{equation}
    h(x)=\frac{\tanh^{-1}(x)}{x},
    \label{FHfunction}
\end{equation}
then equation (\ref{eq:eqatanh}) becomes:
\begin{equation}
    h(z)=1+(h(y)-1)/N.
\end{equation}

We call this function, $h()$, the ``FH function'' since it features prominently in our solution method for the Flory-Huggins model. Using the inverse FH function, $h^{-1}(x)$, which returns the positive branch, we can solve the equation for $z$ in terms of $y$:
\begin{equation}\label{eq:eqZdef}
    z=h^{-1}\left(\frac{1}{N}h(y)-\frac{1}{N}+1\right).
\end{equation}
Therefore, at this point, we have expressed the solution for the phase diagram exactly in terms of $y$ and $z$. The function $h^{-1}(x)$ can be evaluated easily using predefined lookup tables (our strategy), though other strategies based on convergent series representations or asymptotic expressions of $h^{-1}(x)$ are possible. We leave these approaches to future practitioners if they are needed.

From the definition of $y$ and $z$, we can find $\phi_A+\phi_B$:
\begin{equation}\label{eq:eqsumdef}
    \phi_A+\phi_B=\frac{2z}{y+z}.
\end{equation}
Finally, we can return to equation (\ref{eq:eqChemEqChiB}), expressing $\chi$ in terms of $y$:
\begin{equation}\label{eq:eqCHIdef}
    \chi=\frac{\left(\frac{1}{N}-1\right)(\phi_A+\phi_B)y-\ln\left(\frac{1-z}{1+z}\right)}{(\phi_A+\phi_B)^2y},
\end{equation}
where we substitute in for $z$ and $\phi_A+\phi_B$ in terms of $y$ from equations (\ref{eq:eqZdef}) and (\ref{eq:eqsumdef}).

As mapped out in the schematic in Figure~\ref{Figure1-IllustrationOfMethod}a, to construct the binodal curve, we specify a value of $y$ between 0 and 1. Then, we find $z$ by applying equation (\ref{eq:eqZdef}), then $\phi_A+\phi_B$ by applying equation (\ref{eq:eqsumdef}), then the corresponding $\chi$ value by applying equation (\ref{eq:eqCHIdef}). Note that we do not need to explore negative values of $y$ since we arbitrarily assert that the $A$ phase is enriched in the polymer relative to the $B$ phase. 
If we wish to cast the results in terms of $\phi_A$ and $\phi_B$ we can write:
\begin{equation}
    \begin{split}
        &\phi_A=\frac{z(1+y)}{z+y}\\
        &\phi_B=\frac{z(1-y)}{z+y}.
    \end{split}
\end{equation}
If we enumerate multiple values of $y$ between 0 and 1, we can construct the full phase diagram. Our solution is therefore an implicit function, $\chi(y)$, which we can map to $\chi(\phi)$. 

The exact solutions to the Flory-Huggins theory are plotted in Figure~\ref{Figure2PolymerSolventSolution} for $N=$ 10, 50 , 100, and 500. In Figure~\ref{Figure2PolymerSolventSolution}(a-b), we show the functions $z(y)$ and $\chi(y)$ for each value of $N$, which we then map to the function $\chi(\phi)$ in Figure~\ref{Figure2PolymerSolventSolution}c. We confirm that the exact analytical solutions match numerical solutions of the coupled nonlinear equations.  Surprisingly, the approximation first used by Flory~\cite{flory1942thermodynamics} is able to capture the asymptotic behavior of the phase distribution, especially for large $N$.  It seems that this approximation has not received significant attention, since most attention has been paid to analytical approximations in the region near the critical point and extensions from the critical point. For these reasons, we briefly explain the approximation method of Flory, as it shares some similarities to the implicit method we have applied and may be of practical use.

\subsection{Extending Flory's approximation}
Interestingly, in his 1942 article, Flory solved his theory in a rather ingenious way~\cite{flory1942thermodynamics} using an asymptotic arguments of the osmotic pressure difference between solutions. Recognizing that the osmotic pressure of the dilute phase would go to zero relative to the condensed phase, Flory solved the osmotic pressure equation for $\chi$ keeping only terms containing the condensed phase volume fraction. Specifically, he recognized that in the limit $N\gg 1$, and away from the critical point, it is natural to expect that $\phi_B \ll 1$, i.e., the solution is very dilute in polymer, and $\phi_A\approx 1$, i.e., the second phase is rich in polymer. In this limit, it follows that equation (\ref{eq:eqChemEqB}) simplifies to: 
\begin{equation}\label{eq:eqChemEq2}
    \left(\frac{1}{N}-1\right)\phi_A-\ln\left(1-\phi_A\right ) -\chi\phi_A^2=0.
\end{equation}
In other words, there is an implicit relation giving $\phi_A(\chi, N)$ in the form
\begin{equation}\label{eq:eqChemEq3}
    \chi = \frac{\left(\frac{1}{N}-1\right)\phi_A-\ln\left(1-\phi_A\right ) }{\phi_A^2}.
\end{equation}
Flory used this relationship in order to approximate the $\chi$ that corresponds to a given dense phase concentration ($\phi_A$), then iteratively solved for the dilute phase concentration  ($\phi_B$) using the chemical potential constraint.

Interestingly, the starting approximation in equation (\ref{eq:eqChemEq3}) (with $\phi_A\gg \phi_B$) can be applied to the chemical potential equation (\ref{eq:eqChemEqA}) directly to arrive at an implicit approximation for the dilute volume fraction, $\phi_B$, 
\begin{equation}\label{eq:eqChemEq4}
\frac{1}{N} \ln \phi_B  = \frac{1}{N}\ln\phi_A - \ln\left (1-\phi_A\right )-2\chi \phi_A
\end{equation}or 
\begin{equation}\label{eq:eqChemEq5}
\phi_B = \frac{\phi_A}{\left (1-\phi_A\right )^N}e^{-2N\chi \phi_A},
\end{equation}where  $\chi$ for a specified $\phi_A$ is given in equation (\ref{eq:eqChemEq3}).
This last result illustrates that typically $\phi_B$ is exponentially small.

As seen in Figure ~\ref{Figure2PolymerSolventSolution}c, this asymptotic formula works incredibly well to describe the coexisting phases, especially for large $N$, with the exception of a small region near the critical point. Note that the approximation is computed for $2\phi_c<\phi_A<1$, with $\phi_c$ defined by equation \ref{eq:eqCriticalPoint}. Other approximations near the critical point are well documented, and they can also be extended to regions further from the critical point \cite{qian2022analytical, van2023predicting}.

Now that we have solved the single-polymer case, we can turn our attention to what we can learn about multicomponent mixtures.

\section{Multicomponent mixtures}
\label{MulticomponentMixtures}

Most polymer solutions contain more than one polymeric species. Even for solutions containing a polymer of one type, the polymer sample often contains polymers of differing lengths (molecular weight), and each polymer of a specified length must be treated as a separate species~\cite{huggins1967theoretical, van2023theoretically}. 

Here, we will introduce the extension of the Flory-Huggins model to multiple components. Then, the procedure for implementing the implicit substitution method for this system is expounded. Since the number of variables scales with the number of components, inevitably, we must introduce numerous composite variables along the way, which we define as needed. However, the solution strategy remains the same---working with variables $z$ and $y_i$, then using substitution of $\chi$ to arrive at one nonlinear equation with one unknown, and then specifying only the allowed number of composite variables to arrive at a valid solution.

\subsection{Theoretical background }

For a multicomponent system with $M$ polymeric species of volume fraction $\phi_i$ and size $N_i$ in a solvent, the free energy density is ~\cite{mao2019phase}:
\begin{equation}
\begin{split}
    \tilde{f}&=\sum_i \frac{\phi_i}{N_i}\ln(\phi_i)+\left(1-\sum_i \phi_i\right)\ln\left(1-\sum_i \phi_i\right)\\
    &-\chi \sum_i \sum_j \alpha_{ij}\phi_i\phi_j,
    \end{split}
\end{equation}
where the sums are over all $M$ polymeric components (not including the solvent). The incompressibility constraint has been applied to eliminate the solvent density, and linear terms in $\phi_i$ are not included, since they will not influence thermodynamic equilibrium. The coefficients $\alpha_{ij}$ indicate the shape of the symmetric effective interaction matrix, while the prefactor $\chi$ encodes the magnitude of interactions. We assume that the matrix $\pmb{\alpha}$ is known. Up to a linear term, the free energy reduces down to the form  in equation (\ref{eq:free_energy_one_polymer}) if there is only one polymeric specie with $\alpha_{11}=1$.

The chemical potential of each species, $\mu_i=\partial \tilde{f}/\partial \phi_i$, is up to an additive constant:
\begin{equation}
    \mu_i= \frac{1}{N_i}\ln(\phi_i)-\ln\left(1-\sum_j \phi_j\right)-\chi\sum_j 2 \alpha_{ij}\phi_j.
    \label{MulticomponentChemicalPotential}
\end{equation}

The osmotic pressure of the solution is computed as $\Pi=-\tilde{f}+\sum_i \phi_i \partial \tilde{f}/\partial \phi_i$:
\begin{equation}
    \Pi=\sum_i\left(\frac{1}{N_i}-1\right)\phi_i-\ln\left(1-\sum_i \phi_i\right)-\chi \sum_i \sum_j \alpha_{ij}\phi_i\phi_j.
    \label{MulticomponentOsmoticPressure}
\end{equation}

We will assume that two phases $A$ and $B$ in equilibrium have composition $\mathbf{\Phi}_A$ and $\mathbf{\Phi}_B$, where $\mathbf{\Phi}$ indicates a vector with the components consisting of the corresponding chemical volume fractions. These phases satisfy the chemical equilibrium constraints of $\mu_{iA}=\mu_{iB}$ and $\Pi_A=\Pi_B$. Here we will have $M+1$ equations and $2M+1$ unknowns ($\mathbf{\Phi}_A$, $\mathbf{\Phi}_B$, and $\chi$). Therefore, we can set only $M$ variables independently.  

In order to make analytical progress  simplifying the set of nonlinear equilibrium constraints, we next cast the equations into a set of composite variables. The following key composite variables are defined in the next section: $y_i$, the partitioning of component $i$ in equation \ref{eq:eqyidefinition_in}; $z$, the partitioning of solvent in equation \ref{eq:eqzdefinition_in}; $\beta_i$, the sum of the coexisting volume fractions of component $i$ in equations \ref{eq:eqzdefinition_in} and \ref{eq:beta_defined} ; $\gamma_i$ the differential of interactions of component $i$ between phases in equations \ref{MulticomponentGammaI} and \ref{MulticomponentDefinitiongammai}; $\eta_i$ the differential of interactions of component $i$ between phases relative to component 1 defined after equation \ref{eq:masterequation}, and finally $w_i$, the relative partitioning of component $i$ relative to component 1 in equation \ref{eq:wdefinition_in}. A reader who wishes to skip to the final working equations can skip to equations \ref{eq:masterequation} and \ref{eq:eqyidefined}, which together define the simplified master equation in one variable, $z$.

\subsection{Casting into composite variables}

Following the pattern from the one-component case with the implicit substitution method (Section~\ref{OneComponentCase}), we  simplify the arguments of the $\ln$ functions by defining $y_i$ as 
\begin{equation}\label{eq:eqyidefinition_in}
    y_i=\frac{\phi_{iA}-\phi_{iB}}{\phi_{iA}+\phi_{iB}},
\end{equation}
which are between -1 and 1; these variables represent the partitioning of component $i$ between phases. Arbitrarily, we choose $0\leq y_1 <1$, although all other $y_i$ can take on negative or positive values.

The analogous variable for the solvent partitioning is denoted $z$:
\begin{equation}\label{eq:eqzdefinition_in}
    z=\frac{\sum_i \left(\phi_{iA} - \phi_{iB} \right)}{2-\sum_i \left(\phi_{iA}+ \phi_{iB}\right) },
\end{equation}
which again corresponds to the difference divided by the sum of volume fractions of solvent in respective phases, as in equation \ref{eq:eqZsolvent}. If we define the set of variables $\beta_i$,
\begin{equation}\label{eq:eqbetaidefinition_in}
    \beta_i=\phi_{iA}+ \phi_{iB},
\end{equation}
as the sum of volume fraction of component $i$ in coexisting phases, then we can relate $z$ and $y_i$:
\begin{equation}\label{eq:beta_identity}
    \sum_i y_i \beta_i =\left (2-\sum_i \beta_i\right )z.
\end{equation}

To start solving the system analytically, we first solve each constraint for $\chi$. The $M$ chemical potential equations (\ref{MulticomponentChemicalPotential}) give:
\begin{equation}
    \chi=\frac{\frac{1}{N_i}\ln\left(\frac{1+y_i}{1-y_i}\right)-\ln\left(\frac{1-z}{1+z}\right)}{\gamma_i}\quad\hbox{for every $i$},
\end{equation}
with the expression in the denominator, $\gamma_i$, defined as:
\begin{equation}
    \gamma_i=\sum_j 2 \alpha_{ij}\beta_j y_j.
    \label{MulticomponentGammaI}
\end{equation}
Therefore, $\gamma_i$ is a measure of the differential of interactions of component $i$ between phases.

From the osmotic pressure equation (\ref{MulticomponentOsmoticPressure}), we get:
\begin{equation}
    \chi=\frac{\sum_i\left(\frac{1}{N_i}-1\right)\beta_i y_i-\ln\left(\frac{1-z}{1+z}\right)}{\gamma_0}
\end{equation}
with $\gamma_0$ defined as:
\begin{equation}
\begin{split}
    \gamma_0&=\sum_i \sum_j \alpha_{ij}\left(\phi_{iA}\phi_{jA}-\phi_{iB}\phi_{jB}\right)\\
    &=\sum_i \sum_j \alpha_{ij} y_j \beta_i \beta_j=\frac{1}{2}\sum_i \beta_i \gamma_i.
    \end{split}
\end{equation} 

As in Section~\ref{OneComponentCase}, we can  conveniently cast the equations for $\chi$ in terms of $\tanh^{-1}$ functions:
\begin{equation}
\begin{split}
    &\chi=\frac{\frac{2}{N_i}\tanh^{-1}(y_i)+2\tanh^{-1}(z)}{\gamma_i}\quad\hbox{for every $i$},\\
    &\chi=\frac{2\tanh^{-1}(z)+\sum_i\left(\frac{1}{N_i}-1\right)\beta_i y_i}{\gamma_0}.
\end{split}
\end{equation}

For compact notation, we will define the variable $x$ as:
\begin{equation}
    x=\tanh^{-1}(z)
\end{equation}
so that $z=\tanh(x)$, and variables $x_i$  as
\begin{equation}
    x_i=\tanh^{-1}(y_i).
\end{equation}

Before proceeding we can simplify the expressions for $\gamma_i$ (equation~\ref{MulticomponentGammaI}) by defining a variable $w_i$, which is a measure of the relative partitioning of component $i$ relative to the partitioning of component $1$,
\begin{equation}\label{eq:wdefinition_in}
    w_i=\frac{\beta_i y_i}{\beta_1 y_1}=\frac{\phi_{iA}-\phi_{iB}}{\phi_{1A}-\phi_{1B}},
\end{equation}
with $w_1=1$. The definition of $\gamma_i$ in terms of the set $w_i$ is therefore
\begin{equation}
\begin{split}
        \gamma_i&=\beta_1 y_1\sum_j 2 \alpha_{ij}w_j, \quad\quad i\neq 0 \\
    \gamma_0&=\beta_1 y_1\sum_i \sum_j \alpha_{ij} w_j \beta_i.
\end{split}
\label{MulticomponentDefinitiongammai}
\end{equation}

With the definition of $w_i$, by applying equation (\ref{eq:beta_identity}) we can solve for $\beta_i$ in terms of the set of $y_i$, $z$, and $w_i$,
\begin{equation}\label{eq:beta_defined}
    \beta_i=\frac{2 z w_i}{y_i\sum_j\left[w_j\left(1+\frac{z}{y_j}\right)\right]}.
\end{equation}

To summarize the mathematical manipulations thus far, we have recast the equations in terms of composite variables that simplify the arguments of the nonlinear logarithmic functions. Next, we work in terms of these composite variables, until we can write one master equation with one unknown by specifying a maximum of $M$ independent composite variables.

\subsection{Deriving the master equation}

The next step, using our implicit substitution method, is to equate the equations for $\chi$. For this purpose, we  first equate the $\chi$ from the first component to all other components, to obtain $M-1$ identities. Then we  equate a convenient linear combination of $\chi$ from all components to the $\chi$ specified by the osmotic pressure, in order to cancel the denominator $\gamma_0$. Proceeding, 
the combined chemical potential equations give
\begin{equation}\label{eq:eqchempotential_constraints}
        \frac{\frac{2}{N_i}x_i+2x}{\gamma_i}=\frac{\frac{2}{N_1}x_1+2x}{\gamma_1},
\end{equation}
which rearranged is
\begin{equation}
    x_i=x\left(\frac{\gamma_i}{\gamma_1}-1\right)N_i+\frac{\gamma_i}{\gamma_1} \frac{x_1 N_i}{N_1}.
\end{equation}

We  next apply the osmotic pressure constraint on $\chi$, which we write as:
\begin{equation}
    \frac{2x+ \sum_j\left(\frac{1}{N_j}-1\right)\beta_j y_j}{\gamma_0}=\frac{\frac{2}{N_i}x_i+2x}{\gamma_i}.
\end{equation}
Multiplying this equation by $\beta_i\gamma_i$, then summing over all components, we obtain
\begin{equation}
    \frac{\sum_i \beta_i\gamma_i}{\gamma_0} \left(2x+ \sum_j\left(\frac{1}{N_j}-1\right)\beta_j y_j\right)=\sum_i\beta_i\left(\frac{2x_i}{N_i}+2x\right).
\end{equation}
Applying the definition of $\gamma_0$ so as to eliminate it, we find
\begin{equation}
     \left(2x+ \sum_j\left(\frac{1}{N_j}-1\right)\beta_j y_j\right)=\sum_i\beta_i\left(\frac{x_i}{N_i}+x\right).
\end{equation}
Then, collecting terms in $x$, we get:
\begin{equation}
    \left(2-\sum_i\beta_i\right)x-\sum_i\beta_i y_i=\sum_i \frac{\beta_i x_i}{N_i}-\frac{\beta_i y_i}{N_i},
\end{equation}
which can be written compactly as:
\begin{equation}
     \left(2-\sum_i\beta_i\right)(x-z)=\sum_i \frac{\beta_i(x_i-y_i)}{N_i}
\end{equation}
or as:
\begin{equation}
    \frac{\tanh^{-1}(z)}{z}=1+\frac{\sum_i \beta_i(\tanh^{-1}(y_i)-y_i)/N_i}{\sum_i \beta_i y_i},
\end{equation}
where we have reverted back to $y_i$ and $z$ from $x_i$ and $x$. Further, substituting in for $\beta_i$ from equation (\ref{eq:beta_defined}) gives:
\begin{equation}
    \frac{\tanh^{-1}(z)}{z}=1+\frac{\sum_i w_i(\tanh^{-1}(y_i)/y_i-1)/N_i}{\sum_i w_i}.
\end{equation}
We can again write this expression in terms of the FH function, $h()$, defined by equation~(\ref{FHfunction}):
\begin{equation}\label{eq:masterequation}
    h(z)=1+\frac{\sum_i w_i(h(y_i)-1)/N_i}{\sum_i w_i}.
\end{equation}
Defining $\eta_i=\gamma_i/\gamma_1$, which is the differential measure of interactions between phases for component $i$ relative to component 1, the chemical potential constraints in equation (\ref{eq:eqchempotential_constraints}) give:
\begin{equation}
    x_i=N_i(\eta_i-1)x+\eta_i N_ix_1 /N_1
\end{equation}
or in terms of $y_i$:
\begin{equation} \label{eq:eqyidefined}
    y_i=\tanh(N_i(\eta_i-1)\tanh^{-1}(z)+\eta_i N_i\tanh^{-1}(y_1) /N_1).
\end{equation}

At this stage, we have reached our stated goal---to find one equation in one unknown by specifying at most $M$ independent variables. With careful inspection, one may see that specifying $y_1$ and the set $w_i$ for $i>1$ (which constitutes $M$ variables) is sufficient to achieve this objective. Stated more mechanically, by substituting equation (\ref{eq:eqyidefined}) into equation (\ref{eq:masterequation}), and specifying the $M$ composite variables $y_1$ and $w_i$ for $i>1$, we have successfully transformed our system of $M+1$ equations into one master equation, equation (\ref{eq:masterequation}), in one unknown, $z$, the solvent partitioning between phases. 

However, this equation does not have a clear inversion formula, since $z$ is contained differently as an argument of multiple nonlinear functions, not just one nonlinear function as in the one polymer case. Therefore, in the most general case, the master equation must be solved numerically, but there are still significant advantages of this approach compared to the original numerical challenge, as we characterize below. Note however, that there are clear strategies to approximate the master equation solution near and far from the critical point or when a particular pair interaction is dominant, as outlined in Appendices A-D.

\begin{figure*}[t!]
\centerline{\includegraphics[scale=0.45]{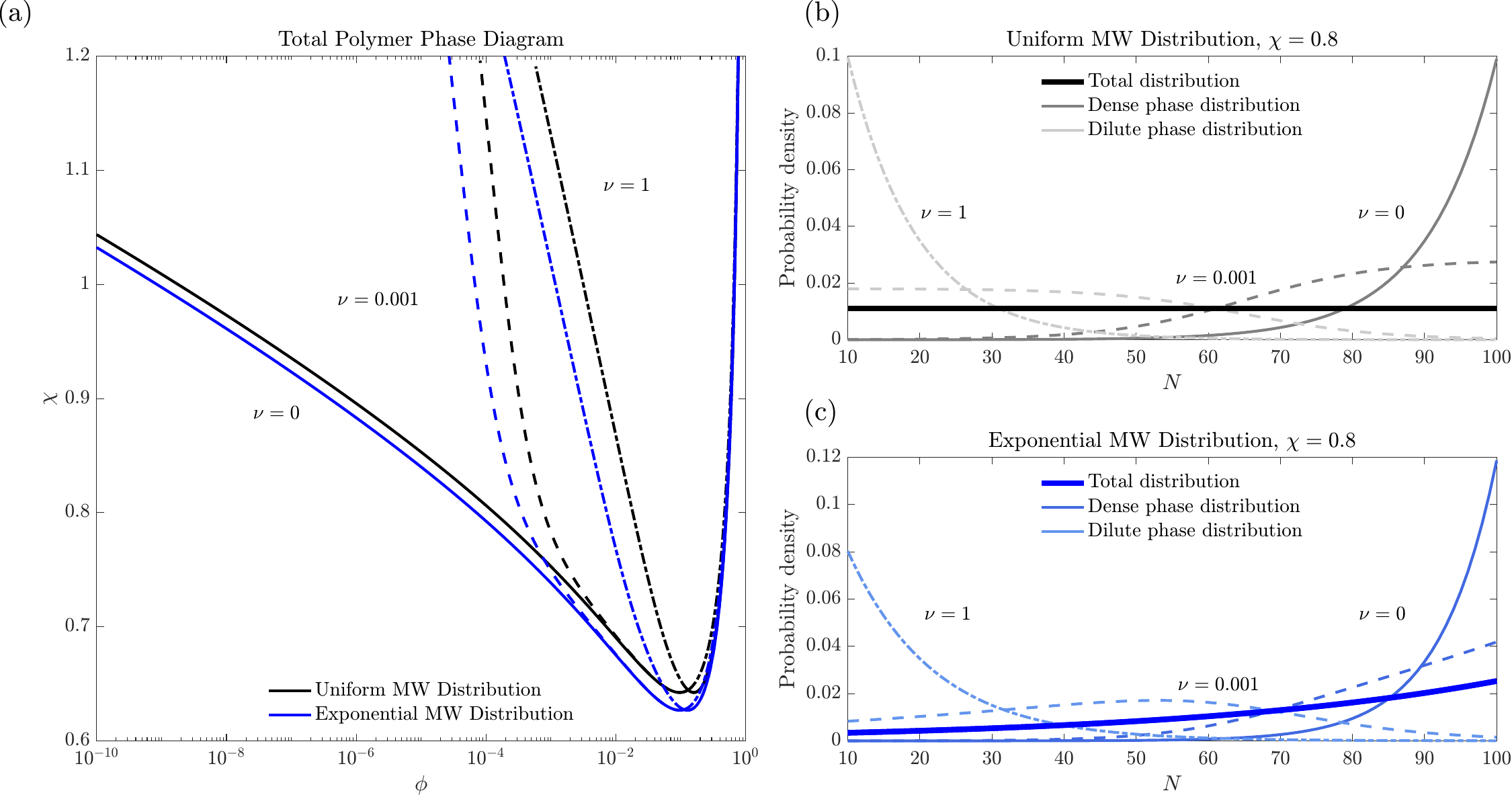}}
\caption{Polydisperse polymer-solvent solution. Comparison of the phase diagram for two polydisperse samples composed of one polymer type. (a) Coexistence curves plotting $\chi$ versus total volume fraction of polymer $\phi=\sum_i \phi_i$. Each curve fixes the volume fraction of the condensed phase, $\nu$. The binodal curves are calculated with a polymer with a uniform molecular weight distribution between $N=10-100$ such that $M=91$, and separately for a polymer with an exponential molecular weight distribution between $N=10-100$, $\bar{\phi}_i/\bar{\phi}_1=\exp((N_i-10)/45)$. (b-c) The molecular weight distributions of the overall solution (thick lines), and the molecular weight distributions in the dense and dilute phases at $\chi=0.8$. Note that for $\nu=0$ and $\nu=1$, the dilute and dense phases, respectively, perfectly overlap with the overall molecular weight distribution.}
\label{FigurePolydispersePolymerSolvent}
\end{figure*}

Next, instead of choosing $w_i$ directly, we choose the set of $w_i y_1=(\phi_{iA}-\phi_{iB})/(\phi_{1A}+\phi_{1B})$ for $i>1$ and $y_1$, so that the specified variables have the same denominator of $\phi_{1A}+\phi_{1B}$, and so that they all go to zero at the global critical point. From $w_i$, one can determine the set $\eta_i$. Then, the master equation can be solved for $z$ numerically. With $z$ in hand, all values of $y_i$ and $\beta_i$ can be calculated explicitly, then finally $\chi$ can be calculated using one of the original equilibrium constraints.

Even though a numerical solution of the master equation is still needed in general, a vast simplification has been achieved by all the algebraic manipulations entailed by the implicit substitution method--- turning the system of $M+1$ nonlinear equations into one nonlinear master equation
(\ref{eq:masterequation}) with one unknown. Again, we must work with the composite variables $y_1$ and $w_i$, but as sketched in the schematic in Figure~\ref{Figure1-IllustrationOfMethod}b-c, the solution can be directly transformed back to $\mathbf{\Phi}$ space with the relation:
\begin{subequations}
    \begin{equation}
        \phi_{iA}= \frac{1}{2}\beta_i(1+y_i)
    \end{equation}
    \begin{equation}
        \phi_{iB}=\frac{1}{2}\beta_i(1-y_i).
    \end{equation}
\end{subequations}
Additionally, we must also check that the solutions returned are actually physical, since admissible solutions to the master equation can return unphysical negative values of the volume fractions.

\subsection{Special cases}

In what follows, we discuss applications of the multicomponent solution method to three special cases. The first case is that of a single polymer with an arbitrary number of different lengths, i.e., the case of polydispersity. There, we find an exact analytical solution, and the master equation solution can be completely cast in terms of two independent variables without any numerical solution necessary. In the second case, we apply the solution method to a two polymer, one solvent system, to demonstrate how the method can be used to find the $\chi$ function over the full compositional space. Finally, in the third case, we suggest strategies to tackle the problem of large multicomponent systems of polymers using the implicit solution without discretizing the entire composition space.

\subsubsection{Polydisperse polymer of one type}
If a solution consists of only a solvent and only one polymer of one type, but that polymer is polydisperse with distinct values of $N_i$, then we can often assume the interactions are equal for each length of polymer, i.e., $\alpha_{ij}=1$, which means that $\eta_i=\gamma_i/\gamma_1=1$ for all $i$. In this case, the value of $y_i$ becomes independent of $z$:
\begin{equation}
    y_i=\tanh\left(\frac{N_i}{N_1}\tanh^{-1}(y_1)\right).
\end{equation}
If this is the case, then we can again solve the equation for $z$ explicitly with the inverse FH function $h^{-1}$, exactly as we did for the one-polymer case:
\begin{equation}
    z=h^{-1}\left(1+\frac{\sum_i w_i(h(y_i)-1)/N_i}{\sum_i w_i}\right).
\end{equation}

For arbitrary molecular weight distributions, the values of $w_i$ and $y_1$ can be chosen independently to construct the full phase diagram. However, in many cases, the molecular weight distribution of the polymer being added to the solvent is known. If the average volume fraction of a particular molecular weight in both phases, $\bar{\phi}_i$, can be written in terms of $\nu$, the volume fraction of the phase $A$ relative to the total volume~\cite{huggins1967theoretical},
\begin{equation}
    \bar{\phi}_i=\nu \phi_{iA}+(1-\nu)\phi_{iB}=(\phi_{iA}-\phi_{iB})\left(\nu+\frac{1}{2 y_i}-\frac{1}{2}\right),
\end{equation}
then the ratio between the average polymer density $\bar{\phi}_i/\bar{\phi}_1$ is:
\begin{equation}
    \frac{\bar{\phi}_i}{\bar{\phi}_1}=\frac{\left(\nu+\frac{1}{2 y_i}-\frac{1}{2}\right)}{\left(\nu+\frac{1}{2 y_1}-\frac{1}{2}\right)}w_i.
\end{equation}
Solving for $w_i$, we get:
\begin{equation}
    w_i=\frac{\bar{\phi}_i}{\bar{\phi}_1}\frac{\left(\nu+\frac{1}{2 y_1}-\frac{1}{2}\right)}{\left(\nu+\frac{1}{2 y_i}-\frac{1}{2}\right)}.
\end{equation}

If $y_1$ and $\nu$ are specified, then the set $w_i$ is also fixed if the overall molecular weight distribution is known. Therefore, we can reduce the dimensionality of the phase diagram by only varying $y_1$ and $\nu$ in the ranges $0<y_1<1$ and $0<\nu<1$, as schematically sketched out in Figure 1b.

In Figure~\ref{FigurePolydispersePolymerSolvent}, we present the phase diagram for two polymer samples of different overall molecular weight distributions, one that is uniformly distributed between $N=10-100$ and one that is exponentially distributed between $N=10-100$, where for both samples the number of species is therefore $M=91$. These curves are analytically calculated in Figure~\ref{FigurePolydispersePolymerSolvent}a, which is much better than working with the 92 nonlinear chemical equilibrium equations we started with. The curves are plotted as $\chi$ versus the overall polymer volume fraction, $\phi=\sum_i \phi_i$, at fixed condensed phase volume fraction, $\nu$. In Figure~\ref{FigurePolydispersePolymerSolvent}b-c, the molecular weight distributions of the samples and the molecular weight distribution of the coexisting phases are plotted in the form of a probability density function. When the dense polymer phase volume fraction is small ($\nu\approx 0$), the binodal is pushed towards lower overall polymer density, because the first appearing condensed phases preferentially include polymers of higher molecular weight, so they dominate the coexistence curve. 

As the volume fraction of the condensed phase, $\nu$, increases, the left branch of the coexistence curve (Figure~\ref{FigurePolydispersePolymerSolvent}) shifts to the right, since the condensed phase is forced to more closely match the overall molecular weight distribution as the condensed phase volume fraction increases. Although $\nu\approx 1$ may be practically infeasible to reach, there, the minority dilute phase must enrich polymers of low molecular weight since they are heavily crowded in the dense phase. Note that at $\nu=0$ and $\nu=1$, the molecular weight distribution of the dilute and dense phase, respectively, perfectly match the molecular weight of the overall polymer sample. Finally, comparing the two overall molecular weight distributions, the sample more enriched in the longer polymers has a left branch of the coexistence curve shifted to the left. In all cases, the critical point appears to be dominated by the largest molecular weight polymer constituents, although it varies with the specific details of the molecular weight distribution and value of $\nu$, as previously found in ~\cite{huggins1967theoretical}.

Also, similar approximations that reduce the compositional dimensionality can be made when one pair interaction is assumed to be dominant, as explained in detail in Appendices C and D, to arrive at another set of analytical solutions.

\begin{figure*}[t!]
\centerline{\includegraphics[scale=0.7]{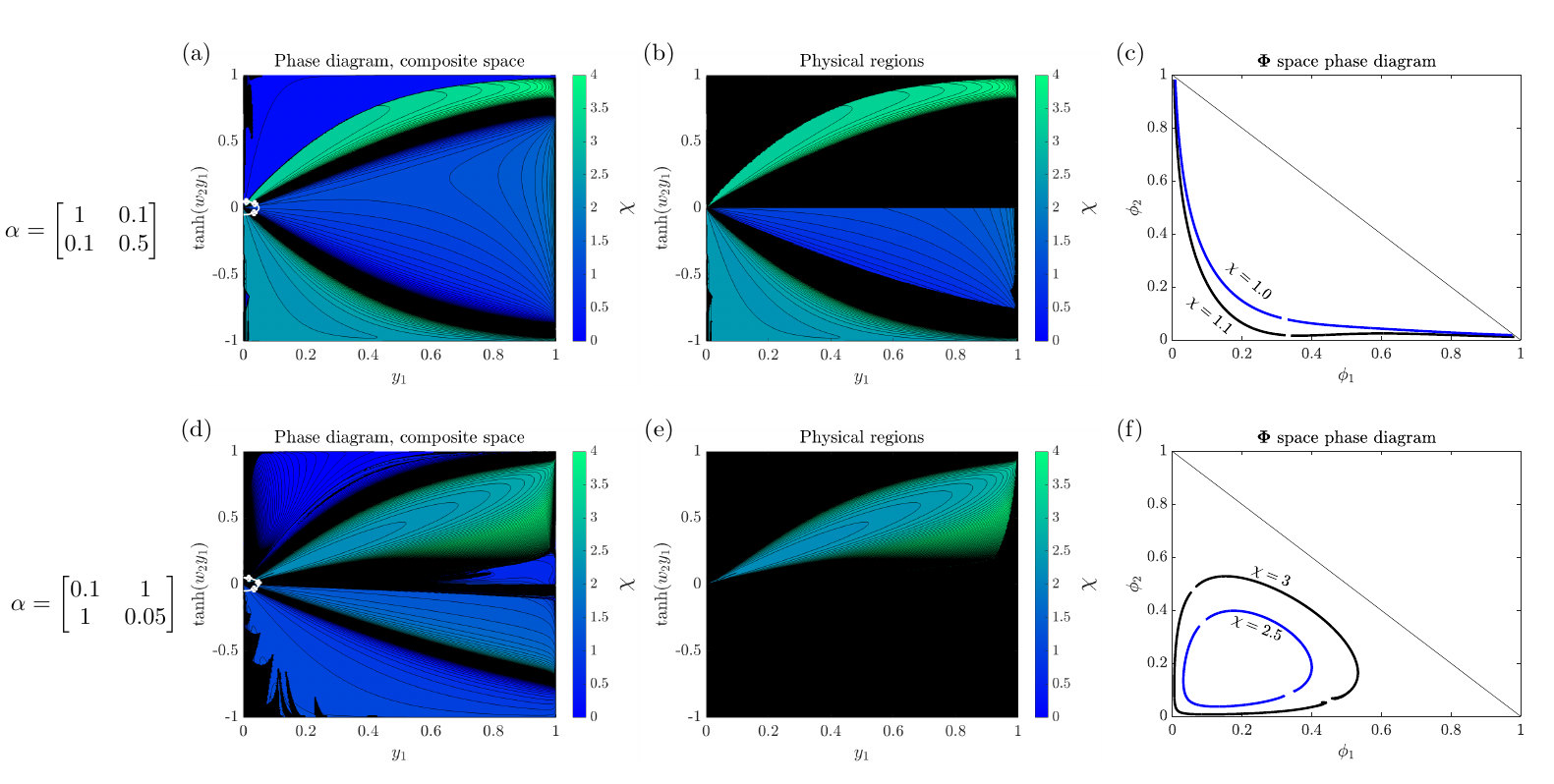}}
\caption{A two-polymer-one-solvent system. Parts (a-c) correspond to the phase diagram for a diagonally dominant interaction matrix, while parts (d-f) show the same results for an off-diagonally dominant interaction matrix. (a) The full phase diagram mapping the $\chi$ surface (with colormap corresponding to the value of $\chi$) for all compositional space for fixed $\alpha_{ij}$ on the $y_1$, $\tanh(y_1 w_2)$ plane. All regions connect at the critical point at $y_1=0$ and $ y_1 w_2=0$, but are segmented by the singular curves on the $\chi$ surface that are evident by thick black regions on the $\chi$ surface. The three poles are computed on a small circle near the critical point, and marked with white markers. These poles can be used to demarcate the distinct differentiable regions of the $\chi$ surface. Note that the colormap is saturated at the bounds of the color bar, but the contour lines are linear between $\chi=-10$ and $\chi=10$ in increments of 0.1, though negative values of $\chi$ do not appear in the feasible physical regions. Black regions correspond to regions where no solution is available or where $|\chi|>10$. (b) The regions of the $\chi$ surface which are physical, with positive predicted polymer densities. (c) The binodal curve for fixed values of $\chi$ transformed to $\mathbf{\Phi}$ space, either mapped out by the gradient descent algorithm in Section~\ref{MulticomponentMixtures}C or by extracting contours from the physical regions of the $\chi$ surface in part 4b. The small gaps correspond to the critical points of the binodal curves, which map to the same point in the composite composition coordinates. (d-f) The same information as (a-c), but for a different interaction matrix. }
\label{Figure4TwoPolymerOneSolvent}
\end{figure*}

\subsubsection{A two polymer, one solvent mixture}
While we were able to find an exact analytical solution for an arbitrarily polydisperse polymer, the two polymer-one solvent mixture has a master equation that can only be solved numerically. Here the master equation becomes:
\begin{equation}
    h(z)=1+\frac{(h(y_1)-1)/N_1+w_2(h(y_2)-1)/N_2}{1+w_2}
\end{equation}
with $y_2$ fixed as:
\begin{equation}
    y_2=\tanh(N_2(\eta_2-1)\tanh^{-1}(z)+\eta_2 N_2\tanh^{-1}(y_1) /N_1)
\end{equation}
and $\eta_2$ defined as:
\begin{equation}
    \eta_2=\frac{\alpha_{12}+\alpha_{22}w_2}{\alpha_{11}+\alpha_{12}w_2}.
\end{equation}

To construct a phase diagram, we specify $y_1$ and $y_1 w_2$, then solve for $\eta_2$, then for $z$ solving the master equation numerically. From there, we can specify $y_2$, $\chi$, and then solve explicitly for the polymer densities in each phase. To discretize the phase diagram, we vary $0<y_1<1$ and $-1<\tanh(y_1 w_2)<1$. Note that it is more regularly behaved to plot the phase diagram in terms of $y_1$ and the product $y_1 w_2$, since they share the same denominator in terms of their $\mathbf{\Phi}$ dependence.

In Figure~\ref{Figure4TwoPolymerOneSolvent}, we present the solution for a two polymer-one solvent system for two different shapes of the interaction matrix: (a-c) a diagonally dominant interaction matrix and (d-f) an off diagonal dominant matrix. So as to not push the coexisting densities to extreme values near 0 or 1, we choose $N_1=4$ and $N_2=3$ as a starting point. The $\chi$ surface is calculated over the full compositional landscape. Interestingly, all differentiable regions emanate from the central critical point at $y_1=0$, $w_2y_1=0$, but are separated by singular discontinuous curves in the phase diagram. Not all of these regions or parts of these regions return physical, positive densities, as shown in Figure~\ref{Figure4TwoPolymerOneSolvent}b,e. Nevertheless, by transforming specified contours with admissible solutions to $\mathbf{\Phi}$ space, we can map out the phase coexistence curves in terms of $\phi_1$ and $\phi_2$, as demonstrated in Figure~\ref{Figure4TwoPolymerOneSolvent}c,f.

Clearly for $M=2$, if the value of the desired $\chi$ contour is known, mapping out the full $\chi$ landscape may be more computationally expensive than solving the equations in $\phi$ space for the fixed, desired value of $\chi$. So the fully discretized $\chi$ surface implemented here are of limited use for generating single binodal curves at fixed $\chi$.

A natural question at this point to ask is: how can one efficiently sample the implicit $\chi$ surface to find a desired contour or sets of contours corresponding to multiple coexisting phases?  Further, how can one choose sampling points so as to dodge the singular curves in the phase diagram? Finally, can the sampling be efficiently scaled for large values of $M$? To explore these questions, in the next section, we formulate such a sampling strategy.

\subsubsection{Finding contours in a high dimensional phase diagram}
In the case of a high dimensional multicomponent solution, we have found an implicit solution for $\chi$, requiring only one nonlinear solve in one variable, $z$, to construct the surface:
\begin{equation}
    \chi=\chi(y_1, \tanh(y_1 w_i))=\chi(\mathbf{v}),
\end{equation}
where only the $M$ independent variables are listed as arguments in the vector $\mathbf{v}$.

Without enumerating all possible inputs for the implicit function $\chi$, we may not know beforehand where a given contour lies on the $\chi$ surface corresponding to coexisting phases. If we wish to only find the contours of this surface at fixed $\chi=\chi_0$ with minimal number of $\chi$ evaluations in a high dimensional phase diagram ($M\gg 1$), then we can employ a gradient descent algorithm on the $\chi$ surface to arrive at the desired contour. Defining the inputs of the $\chi$ function as a vector $\mathbf{v}$, with step index $k$, the algorithm proceeds as:
\begin{equation}
    \mathbf{v}_{k+1}=\mathbf{v}_k-\lambda_k \nabla_v F(\mathbf{v}).
\end{equation}
The function $F$ is:
\begin{equation}
    F(\mathbf{v})=\left(\chi(\mathbf{v})-\chi_0\right)^2,
\end{equation}
and its gradient can be computed by finite difference. The step size $\lambda_k$ can be specified by a given algorithm, such as the Barzilai-Borwein method~\cite{barzilai1988two}. 


Once the desired contour is reached, one can then trace out a contour by stepping along a direction, $\mathbf{u}$, perpendicular to the gradient $\mathbf{g}=\nabla_v \chi$. Note, however, that there are $M-1$ such coordinate directions that are perpendicular to $\mathbf{g}$. Here, we choose just one of these coordinate directions,
\begin{equation}
\mathbf{v}_{k+1}=\mathbf{v}_k-\lambda_k \mathbf{u}
\end{equation}
with 
\begin{equation}
\begin{split}
        &u_i=g_i, \quad \quad i>1 \\
        &u_i=-\frac{1}{g_1}\sum_{i>1} g_i^2 \quad \quad i=1.
\end{split}
\end{equation}

Here the step size $\lambda_k$ can be scaled by the magnitude of the gradient vector or the location of the current iteration, $\mathbf{v}_k$. The direction of the step away or towards the global critical point can be controlled by the sign of $\lambda_k$. If ever the step size is too large and the algorithm steps off of the fixed $\chi$ contour, the gradient descent can be used to return to the contour.  

In a high dimensional composition space, the binodal curves become higher dimensional surfaces. Stepping in one tangential coordinate direction may not sample the entire binodal surface on its own. Therefore, one would need to choose a path that efficiently maps the boundary of the binodal surface, and chooses the tangential directions based on the distance from the critical point, the attributes of the gradient vector, and the historical location of points already sampled. Here, we do not derive such a high dimensional search algorithm.

The gradient descent algorithm works to converge to one contour in the $\mathbf{v}$ space, corresponding to phase coexistence points in the $\mathbf{\Phi}$ space. If the algorithm is initiated in a region that is separated from the desired contour by a singular, non-differentiable curve in the $\mathbf{v}$ space, then more than one starting point would be needed to locate all matching contours in the phase diagram. If this is the case, then it is natural to ask: what are efficient ways to choose candidate starting points?

Notably, all regions connect at the global critical point at $\mathbf{v}=0$, but the regions are separated by singularities which extend radially from the critical point, slicing the phase diagram like a pie, as seen clearly in Figures 4a and 4d. Therefore, the analytical approximations of the equations by expanding near the critical point and finding poles can give a small number of candidate starting points in each region located at the center of each pie piece.

To sketch out the strategy to initiate a gradient descent algorithm, we next turn to how to segment the phase diagram efficiently to generate candidate starting points near the critical point.

\textit{{Segmenting a hypersphere near the critical point}}:

Near the critical point, the partitioning of each species becomes negligible, i.e., $|y_i|, |z| \ll 1 $, and the master equation is:
\begin{equation}\label{eq:eqCrit}
    z^2=\frac{\sum_i w_i y_i^2/N_i}{\sum_i w_i}
\end{equation}
with the chemical potential constraint setting:
\begin{equation}
    y_i=N_i(\eta_i-1)z+\eta_i \frac{N_i}{N_1} y_1.
\end{equation}
Combined, these equations give a quadratic equation for $z$ which can be solved exactly, as documented in Appendix B. Nevertheless, extra degrees of freedom are still specified by the set $y_1$, $w_i$. 

To eliminate these degrees of freedom, we propose the following strategy. We look for candidate starting points on a small hypersphere $| \mathbf{v} |=v_0$ near the critical point, with $v_0\ll 1/\max(N_i)$. First, we find the poles of the $\chi$ function on this hypersphere. Then, we choose a small number of points in regions demarcated by the poles of the $\chi$ function. By choosing specified points between the poles, we guarantee that we are testing each candidate region for the specified $\chi$ value. 

For finite but small $z$ and $y_i$, the poles of the $\chi$ function are defined by the equation:
\begin{equation}
    \gamma_i\propto\sum_j \alpha_{ij} w_j=0 
\end{equation}
which corresponds to finding the null space of the $\mathbf{\alpha}$ matrix, and then rescaling so that $w_1=1$. If $\mathbf{\alpha}$ is full rank, then the null space is empty. However, one other denominator in the full expressions can go to zero, causing singularities in the phase diagram, namely:
\begin{equation}
    \sum_j\left[w_j\left(1+\frac{z}{y_j}\right)\right]=0.
\end{equation}
Near the critical point, these poles must be solved for numerically for the values of $w_i$ and $y_1$ that lead to singularities, making use of the exact solution for $z$ near the critical point. 

Note that the poles can actually be interpreted as curves on a given hypersphere. For $M=2$, they correspond to points on a circle. For $M=3$, they correspond to arcs on a sphere, and so on. In order to segment the space in higher dimensions, one would need to find the intersection of these singular curves on the hypersphere near the critical point, to find the boundaries of the nonsingular ``continents'' on the hypersphere.

Once we have found the poles near the critical point at fixed $v_0$, we can segment the hypersphere radially between these poles, making sure to test at least one point in the gradient descent algorithm in each ``continent" of the segmented hypersphere. This would ensure that our algorithm can find a desired contour, if it exists, and could find multiple contours of the same function that are separated by non-differentiable curves.

\textit{{Applying the gradient descent algorithm}}:

Here, we explicitly apply the gradient descent algorithm only for $M=2$, to find the contours in Figures~\ref{Figure4TwoPolymerOneSolvent}c,f, validating the results by comparison to the full $\chi$ surface. The poles of the function do segment the phase diagram for $M=2$, as demonstrated by the white points marked in Figures~\ref{Figure4TwoPolymerOneSolvent}a,d on a circle a fixed distance from the critical point. Choosing a handful of starting values between the singular points on the circle, we can trace out the fixed values of $\chi$, and then transform to $\mathbf{\Phi}$ space, to identify curves matching Figures~\ref{Figure4TwoPolymerOneSolvent}c,f.

While we demonstrate the gradient descent algorithm for $M=2$, we leave the exploration of $M>2$ to future work. Note that this gradient descent algorithm is very advantageous for cases where $M>4$, where it may be difficult to discretize the $\mathbf{v}$ space over the whole compositional range a priori~\cite{mao2019phase}. In fact, while the number of function evaluations at each point scales with $M$ (to evaluate the components of the gradient vectors), and the number of candidate points may also increase with $M$, the gradient descent algorithm is otherwise insensitive to $M$, which is much more favorable than discretizing all combinatoric possibilities in composition space. Still, the stepping along the binodal surfaces in high dimensions must resolve enough points so as to map out the coexistance surface to significant resolution, which may be a computationally demanding task.



\section{Conclusions}
Using the implicit substitution method, we are able to analytically define the coexistence curves for a polymer-solvent system and a polydisperse polymer-solvent system.  For generalized mixtures of many components, we have simplified a large system of nonlinear equations into one nonlinear equation in one composite variable. We have presented methods to approximate the solutions, not only for the single-polymer solvent system (Section~\ref{OneComponentCase}C), but also in the case of multicomponent solutions far from their critical point (Appendix A) and close to their critical point (Appendix B). More analytical solutions are possible in arbitrary mixtures if there is one dominant interaction, where all other non-interacting species operate as non-interacting crowders (Appendices C-D).

As a test of the multicomponent master equation, we have mapped out the implicit $\chi$ surface for two polymeric components over the full compositional space, and demonstrated the demarcation between differentiable regions of the $\chi$ surface that defines the binodal curve. Further, we have suggested strategies to find full binodal curves at fixed $\chi$ without discretizing the full compositional space, which we believe will be very useful for $M>4$, but we leave a full exploration of these methods to future studies. 

Ample opportunities exist to develop the methods presented here further. Beyond improving the computational performance of the method for many components, one could seek analytical perturbative series solutions to the general master equation. Further, one could implement the method to large multicomponent mixtures, and explore the shape of contours of the $\chi$ surface for different interaction matrices. Moreover, the approach can be directly extended to more detailed equations of state beyond Flory-Huggins, although compact analytical solutions of a final master equation are not guaranteed for all equations of state.  In Appendix E, we highlight possible extensions of the implicit substitution method to other equations of state.

The composite composition variables are experimentally accessible if the composition of the dilute and concentrated phases are simultaneously measured. Therefore, experimental phase diagrams may also be mapped to composite composition space for direct comparison to the theoretical $\chi$ surface for a particular $\pmb{\alpha}$ matrix at fixed temperature.


The implicit substitution method, while possibly un-intuitive at first sight, has given us a new lens to explore the phase behavior of polymeric phase separation. Here we demonstrated its power in simplifying systems of coupled nonlinear thermodynamic equations that are often impossible to solve explicitly. 

\section{Acknowledgements}
We acknowledge support from the Princeton Center for Complex Materials, an NSF MRSEC (DMR-2011750), and NSF for grant DMS/NIGMS 2245850. J.P.D. is also supported by the Omenn-Darling Princeton Bioengineering Institute -- Innovators (PBI2) Postdoctoral Fellowship. We would like to thank Hongbo Zhao, Tejas Dethe, and Qiwei Yu for helpful discussions and comments.

\appendix
\section{Approximations far from the critical point for multicomponent mixtures}

For multicomponent mixtures, it may be difficult to seek approximations, since multiple assumptions of limiting values of densities may need to be made simultaneously. While the limiting solution near the critical point is analytical   (by applying the quadratic formula in equation (\ref{eq:eqCrit}) as detailed in Appendix B), one general approximation does not work in all regions far from the critical point. 

However, we can explore the limiting behavior when the major component, which we arbitrarily specify as component 1, has a large excess in the condensed phase compared to the dilute phase relative to other components. We also choose a secondary component that exhibits the opposite partitioning compared to component 1, but exhibits more partitioning compare to the other species. Therefore, this boundary corresponds to the edge of the phase diagram in our chosen coordinates, where $y_1 \rightarrow 1$, $w_iy_1\rightarrow \pm \infty$, or $z \rightarrow 1$. 

In our first analysis, we start with the solvent as the secondary dominant component. If this is the case, $y_1\rightarrow 1-\epsilon$, where $\epsilon$ is vanishingly small, with a solvent dominated dilute phase, meaning also $z\rightarrow 1$. We can effectively solve the system as if no other polymers are present. In the limit of $\epsilon\rightarrow 0$, where correspondingly $w_i$ for $i>1$ go to zero, we get
\begin{equation}
    h(z)\approx 1+{(h(y_1)-1)/N_1},
\end{equation}
which has solutions in terms of the inverse FH function:
\begin{equation} \label{eq:eqAPPSolution}
    z=h^{-1}\left(1+{(h(y_1)-1)/N_1}\right).
\end{equation}
We can subsequently solve for the set $y_i$ describing the partitioning of other polymers by applying equation (\ref{eq:eqyidefined}). We can map this approximation to all of composition space for a specific value of $y_1$, $w_i$,  and then check that the master equation is satisfied up to some tolerance to accept the approximation prediction, or to reject it.

In many cases, though, the dilute phase in one polymer is the dense phase in the other, and vice versa, for a given polymer pair. In other words, the solvent may not be significantly partitioned between phases. If that is the case, we can seek approximations by assuming that one of the polymers is the ``solvent" while the other is the polymer, effectively rescaling the theory in these dominant composition coordinates. If we were to do this with component $j$ as the solvent, and component $y_i$ the dominant partitioning component, we would find:
\begin{equation} 
    y_j=-h^{-1}\left(1+{(h(y_i)-1)N_j/N_i}\right).
\end{equation}
where $y_j$ is chosen to have the opposite sign of $y_i$. All other composite variables, $y_k$, can be computed by the analogue of equation (\ref{eq:eqyidefined}) that treats component $j$ as the ``solvent'' component that satisfies incompressibility, namely:
\begin{equation} 
    y_k=\tanh\left(-\frac{N_k}{N_j}\left(\frac{\eta_k}{\eta_i}-1\right)\tanh^{-1}(y_j)+\frac{\eta_k N_k}{\eta_i N_i} \tanh^{-1}(y_i) \right)
\end{equation}
where for the solvent $y_0=-z$.

We can use this procedure to map out the areas near the boundary of our composition space, where $y_1 \rightarrow 1$ or $w_iy_1\rightarrow \pm \infty$. Further, by considering each possible pair of polymers or solvent combination, we can generate a combinatoric set of approximations that span all regions far from the critical point, that may at least make for useful guesses for the full master equation. Similar to the one-polymer case, we expect that these approximations will be most accurate when there are significant differences in the polymer lengths.

\section{Approximations near the critical point}
Near the critical point, the partitioning of each species becomes negligible, i.e., $| y_i|, | z| \ll 1 $, and the master equation is
\begin{equation}
    z^2=\frac{\sum_i w_i y_i^2/N_i}{\sum_i w_i},
\end{equation}
with the chemical potential constraint setting:
\begin{equation}
    y_i=N_i(\eta_i-1)z+\eta_i \frac{N_i}{N_1} y_1.
\end{equation}
This approximation is valid as long as:
\begin{equation}
\begin{split}
        &z\ll 1 \quad \& \quad z\ll \frac{1}{N_i(\eta_i-1)} \\
        &y_1\ll 1 \quad \& \quad y_1\ll \frac{N_1}{N_i\eta_i}.
\end{split}
\end{equation}
 Combined, these equations give a quadratic equation for $z$ which can be solved exactly. The quadratic equation explicitly is: 
\begin{equation}
\begin{split}
        0=&z^2\left(-1+\frac{\sum_i w_i N_i(\eta_i-1)^2}{\sum_i w_i}\right)\\
        &+z\left(\frac{\sum_i 2w_i N_i\eta_i(\eta_i-1)y_1/N_1}{\sum_i w_i}\right)\\
        &+\frac{\sum_i w_i N_i\eta_i^2y_1^2/N_1^2}{\sum_i w_i}.
\end{split}
\end{equation}
Picking the set $w_i$ and $y_1$, we solve for $z$, choosing the root that gives positive $\beta_i$ for all $i$, if in fact either root gives positive $\beta_i$.





\section{Adding non-interacting ``crowders''}
To add to our list of analytical solutions, we can consider the case of non-interacting  crowders added to a single strongly interacting polymer in solvent. Although this ideal non-interacting assumption may be rare in reality, it may be a reasonable approximation if one polymer self-interaction is much larger in magnitude than all other interactions. We assume that the dominant polymer is $i=1$, where $\alpha_{11}=1$.  If non-interacting crowders are added with species index: $2<i<M$, then since their corresponding rows and columns $\alpha_{ij}=0$, we have $\eta_i=0$ for these species. Therefore, the $\pmb{\alpha}$ matrix only has one entry.   

For the non-interacting polymers, we have:
\begin{equation}
    y_i=-\tanh\left(N_i\tanh^{-1}(z)\right).
\end{equation}

The master equation can be written as:
\begin{equation}
    \begin{split}
        h(z)-1=&\frac{\left(h(y_1)-1\right)/N_1}{\sum_i w_i}\\
        &+\frac{\sum_{i=2}^{M} w_i\left(\frac{N_i \tanh^{-1}(z)}{\tanh\left(N_i\tanh^{-1}(z)\right)}-1\right)/N_i}{\sum_i w_i}
    \end{split}
\end{equation}
While $z$ is embedded in multiple nonlinear functions, $y_1$ appears as only one argument of a nonlinear function. Therefore, instead of specifying $y_1$ and solving for $z$, we can specify $z$ and solve for $y_1$:
\begin{equation}
\begin{split}
     h(y_1)=&1+N_1\sum_i w_i \left(h(z)-1\right) \\
     &+\sum_{i=2}^{M} w_i\left(1-\frac{N_i \tanh^{-1}(z)}{\tanh\left(N_i\tanh^{-1}(z)\right)}\right)\frac{N_1}{N_i}.
\end{split}
\end{equation}
In this case, $y_1$ can explicitly be expressed in terms of the inverse FH function.

Similar to the polydisperse problem, we may know the overall ratios of the crowders being added to the solution relative to the total solvent being added. We can express this ratio in terms of $z$ and $\nu$ (the volume fraction of the dense phase A):
\begin{equation}
\begin{split}
        \bar{\phi}_i&=\nu \phi_{iA}+(1-\nu)\phi_{iB}=(\phi_{iA}-\phi_{iB})\left(\nu+\frac{1}{2 y_i}-\frac{1}{2}\right)\\
        &=(\phi_{iA}-\phi_{iB})\left(\nu-\frac{1}{2\tanh(N_i\tanh^{-1}(z)) }-\frac{1}{2}\right)
\end{split}
\end{equation}
then the ratio between the average polymer density of component $i$ and the solvent $\bar{\phi}_i/(1-\sum_j\bar{\phi}_j)$ is:
\begin{equation}
    \frac{\bar{\phi}_i}{1-\sum_j\bar{\phi}_j}=\frac{\left(\nu-\frac{1}{2\tanh(N_i\tanh^{-1}(z)) }-\frac{1}{2}\right)}{\left(-\nu+\frac{1}{2 z}+\frac{1}{2}\right)}\frac{w_i}{\sum_j w_j}.
\end{equation}
If the left hand side of the equation (the ratio the crowder to solvent), is known for $i>1$, and $\nu$ and $z$ are specified, then the above equations give $M-1$ linear equations for the $M-1$ values of $w_i$ for $i>1$.

Therefore, if $\nu$ and $z$ are specified with a known total composition of the crowders and solvent, then the set $y_1$ and $w_i$ are all specified. To construct a phase diagram with crowders present, we can simply vary $\nu$ and $z$ between 0 and 1 to construct the full $\chi$ surface, then transform to $\mathbf{\Phi}$ space.

Note that we may also use this ``non-interacting" construction to include finite ``compressibility" of our solution, by including ``holes" on the lattice as one of the non-interacting species, along with the solvent and polymer species. However, this lattice based approximation is typically a poor approximation for the compressibility of hard-sphere liquids~\cite{frydel2012close}, and therefore can likely only approximate the behavior of real liquids.

\section{ One dominant polymer-polymer interaction}
Analogous to the case of one dominant self-interacting polymer is the case of one dominant polymer-polymer pair interaction, meaning an interaction matrix that is zero everywhere except for two off-diagonal entries. If we are to reframe the theory such that one of the polymers is made into the ``solvent" species, and the other is made into the ``dominant" interacting component, then the approximation becomes equivalent to the previous section, where all other species are only non-interacting crowders that are non-participating observers to the main interacting components. In these reframed variables, the interaction matrix would have only one non-zero entry, $\alpha_{11}$. 

Note that for the case where the ``solvent" species has length $N_0>1$, we can rewrite the master equation as
\begin{equation}
    h(z)=1+\frac{\sum_i w_i(h(y_i)-1)N_0/N_i}{\sum_i w_i}
\end{equation}
and the chemical potential constraints give $y_i$:
\begin{equation}
    y_i=\tanh(N_i(\eta_i-1)\tanh^{-1}(z)/N_0+\eta_i N_i\tanh^{-1}(y_1) /N_1).
\end{equation}
Next, we could pursue the same approximations as in the previous sections where $\eta_i=0$ for $i>1$, then frame the phase diagram in terms of $\nu$ and $z$. We leave an exploration of these approximations to future work.

It should be noted that there is a subtle difference between the ``dominant-interaction-approximation" in this section as compared to the ``far from critical'' approximation in Appendix A. The dominant-interaction approximation is valid only when the set $\eta_i\ll1$ for $i\neq 1$, regardless of the geometric distance from the critical point. However, for sufficiently strong partitioning of a weakly interacting component $k$, $| w_k |\gg1$, the condition on $\eta_k$ will eventually break. So, in other words, the ``dominant-interaction approximation'' breaks very far from criticality for weakly interacting components. The ``far from critical" approximation, on the other hand, is even valid when the interaction matrix has many non-zero entries, though its validity is only maintained far from the critical point. 

To summarize all the approximations explored, building from Flory's original approach for a polymer-solvent system, we have derived an analytical approximation far from the critical point. Next, we have derived an approximation valid in a small region near the critical point. To bridge these approximations, we have explored regions of the phase diagram where a particular polymer-polymer or polymer-solvent interaction is dominant. While more intricate approximations may be sought for triplets of species, we suspect that for most regions in the phase diagram, the phase composition is dominated by the two main interacting accumulated/depleted components between the phases, and if not, we are either close to the global critical point or very far from the global critical point---where other approximations are more readily accessible.

\section{Extensions to other equations of state}
Here, we can explore how this formulation could be extended to other equations of state based on a lattice, focusing only on the polymer-solvent system. In other theories, one could replace the $-\chi \phi^2$ term in the free energy density by:
\begin{equation}
    -\chi \tilde{f}_\mathrm{int}(\phi)
\end{equation}
which is equivalent to assuming a $\phi$-dependent $\chi$ parameter.

In this case, from the chemical potential, we arrive at:
\begin{equation}
    \chi=\frac{\frac{2}{N}\tanh^{-1}(y)+2\tanh^{-1}(z)}{\tilde{f}_\mathrm{int}^{\prime}(\phi_A)-\tilde{f}_\mathrm{int}^{\prime}(\phi_B)}
\end{equation}
And from the pressure, we have:
\begin{equation}
    \chi=\frac{\left(\frac{1}{N}-1\right)(\phi_A+\phi_B) y+2\tanh^{-1}(z)}{\tilde{f}_\mathrm{int}(\phi_B)-\tilde{f}_\mathrm{int}(\phi_A)+\phi_A\tilde{f}_\mathrm{int}^{\prime}(\phi_A)-\phi_B\tilde{f}_\mathrm{int}^{\prime}(\phi_B)}
\end{equation}

Substituting in:
\begin{equation}
    \begin{split}
        &\phi_A=\frac{z(1+y)}{z+y}\\
        &\phi_B=\frac{z(1-y)}{z+y}
    \end{split}
\end{equation}
and equating the values of $\chi$, we generally have a nonlinear equation for one unknown ($z$) in terms of one specified variable ($y$)---reducing our system of two equations down to one equation. While not guaranteed, one may be able to decouple the $z$ and $y$ dependence, defining a function similar to the FH function $h()$ for a given functional interaction form. While this procedure is marginally better than solving the full system of two equations for the polymer-solvent system, it may be especially helpful for the multicomponent case. We leave these generalizations to future work.

\bibliography{library.bib}

\end{document}